# Building Conflict Uncertainty into Electricity Planning: A South Sudan Case Study

Neha Patankar[a], Anderson Rodrigo de Queiroz[b], Joseph F. DeCarolis[c,*], Morgan D. Bazilian[d], Debabrata Chattopadhyay[e]

[a] Operations Research Department, NC State University, Raleigh, NC 27695, United States
[b] Department of Decision Sciences, School of Business, NC Central University, Durham, NC 27707, United States
[c] Civil, Construction, and Environmental Engineering, NC State University, Raleigh, NC 27695, United States
[d] Payne Institute, Colorado School of Mines, Golden, CO, 80401, United States
[e] World Bank, Washington, DC, United States
[*] Corresponding author. E-mail address: jdecarolis@ncsu.edu

**Abstract**

This paper explores electricity planning strategies in South Sudan under future conflict uncertainty. A stochastic energy system optimization model that explicitly considers the possibility of armed conflict leading to electric power generator damage is presented. Strategies that hedge against future conflict have the greatest economic value in moderate conflict-related damage scenarios by avoiding expensive near-term investments in infrastructure that may be subsequently damaged. Model results show that solar photovoltaics can play a critical role in South Sudan's future electric power system. In addition to mitigating greenhouse gas emissions and increasing access to electricity, this analysis suggests that solar can be used to hedge against economic losses incurred by conflict. While this analysis focuses on South Sudan, the analytical framework can be applied to other conflict-prone countries.

**Keywords**: Stochastic programming; conflict uncertainty; South Sudan

## 1. Introduction

Electricity supply security is critically important, especially in fragile and conflict-affected states where resumption of electricity services can restore confidence in the government and society, strengthen security, and revive the economy (World Bank, 2013). Addressing fragility, conflict, and violence (FCV) is required to end poverty and promote shared prosperity (World Bank, 2015). While the provision of affordable and reliable electricity supply can promote economic development and help countries exit the conflict trap (Collier, 2003), electric power systems are also vulnerable to conflict conditions. Attackers in many conflict environments have targeted electricity transmission lines and power generation plants, which can lead to long outages and the need for system restoration (Zerriffi, H. et. al., 2002).

Acknowledging that each conflict has its own unique dynamics (Goldstone, 2008), recommendations should be based on a thorough examination of specific conflict situations. In this paper, we explore potential electricity development pathways in South Sudan. South Sudan has been ranked as the most fragile country in the world for the last several years (Fund for Peace, 2017), and it is also one of the least developed countries in the world. There are approximately 250 km of paved roads and less than 30 MW of installed electric generating capacity serving 13 million people in a landlocked area slightly smaller than the US state of Texas (CIA, 2018). Soon after South Sudan gained its independence in 2011, the government started to attract investment funding for hydropower installations (IEA, 2014). Two years later, in 2013, a civil war erupted and it is still ongoing despite a peace agreement signed in 2015 (The Guardian, 2016). To the best of our understanding, most of the investments in the electricity infrastructure expansion have been suspended. Despite having an abundance of natural resources, conflict in South Sudan makes the country prone to economic collapse (World Bank, 2016).

Electrification strategies under FCV conditions should explicitly consider the risk of conflict in the decision making process. However, this is often not the case. For example, EAPP (2011) examined future electricity development by employing a conventional least-cost capacity planning model and concluded that South Sudan should focus on developing a series of large-scale hydroelectric dams along the White Nile. Political issues were considered, but only exogenously to the optimization model. Such a focus on large scale infrastructure projects with long construction times produces inefficient outcomes. These hydroelectric projects never broke ground, and instead hundreds of millions of dollars have been spent on generators and diesel fuel (Mozersky and Kammen, 2018). While incorporating conflict risk in energy system planning is challenging and subject to considerable uncertainty, it should not be ignored (Bazilian and Chattopadhyay, 2016).

This paper focuses on developing planning strategies for the South Sudan electric power system that explicitly consider conflict uncertainty. We model the South Sudan system using an open source energy system optimization model, and incorporate conflict by performing multi-stage stochastic optimization (Birge and Louveaux, 2011; Pereira and Pinto, 1991; de Queiroz, 2016). Optimization is performed over a scenario tree that represents different conflict-related outcomes in the future, and the resultant stochastic solution suggests a near-term planning strategy. Given the paucity of data and large future uncertainties, we perform sensitivity analysis to identify critical assumptions and develop insights that explicitly consider conflict-related uncertainty.

While the application of stochastic optimization yields a planning strategy, this analysis should nonetheless be viewed as an exercise to explore the decision space when conflict is explicitly considered. We are not able to capture all of the real-world conflict dynamics and potential power system failure modes. In addition, we emphasize that models alone cannot provide a solution in such complex decision landscapes, but can yield insight that informs decision making. This paper is intended to further the discussion between modelers and the decision makers, planners, and consultants who develop electrification strategies in FCV countries.

The results presented here suggest promise for further application. Much of the analysis focused on energy development in Africa has been focused on universal access and climate change mitigation through renewables deployment (Lucas et. al., 2017; Africa Progress Report, 2015; AREI, 2017; Wu et. al., 2017; Deichmann et. al., 2011). Considering conflict-related uncertainty can add another dimension to future analysis, ensuring that energy supply is also resilient in the face of conflict, fragility, and violence.

   2

## 2. Methods

Key aspects of the modeling effort are described in this section. We begin by describing Tools for Energy Model Optimization and Analysis (Temoa), the open source energy system optimization model and the South Sudan input dataset used to conduct this work. Next, we describe Method of Morris, a sensitivity analysis technique that allows us to identify the input parameters with the largest effect on total system cost. Then we describe the stochastic model formulation, the method by which generator damage is estimated, and the metrics used to assess the cost of conflict uncertainty. The appendix provides additional detail on technology specifications, demand projection, and the estimation of damages.

### 2.1 Tools for Energy Model Optimization and Analysis (Temoa)

Tools for Energy Model Optimization and Analysis (Temoa) is an open source, Python-based framework to conduct energy systems analysis. The core component of Temoa is a bottom-up, technology rich energy system optimization model (ESOM). The Temoa model formulation is similar to the MARKAL/TIMES model generators (Fishbone and Abilock, 1981), MESSAGE (Messner and Strubegger, 1995), and OSeMOSYS (Howells et al., 2011). Technologies are represented by a set of engineering-economic parameters, and linked together in an energy system network through a user-specified series of commodity flows. The model employs linear optimization to minimize the system-wide cost of energy supply over the user-defined time horizon by optimizing the installed capacity and utilization of energy technologies. Several constraints ensure appropriate system performance, including energy supply sufficient to meet demand, energy balance at both the process and system-wide levels, and operating limits on baseload plants. The complete algebraic formulation of Temoa is published (Hunter et. al., 2013), and the model source code is publicly available through a GitHub repository (TemoaProject, 2018).

### 2.2 Input Data

A Temoa-compatible dataset to represent South Sudan was created for this analysis. The model time horizon extends from 2017 to 2037, with five-year time periods defined at 2017, 2022, 2027, and 2032. When performing stochastic optimization, conflict uncertainty is resolved in the latter three time stages. The climate of South Sudan is tropical and has a wet and dry season. Most rainfall occurs from May to October while December, January, and February are the driest months. To capture diurnal variation in electricity production from solar PV, each day is split into day and night, and to represent the tropical climate of South Sudan, each year is split in two seasons: wet (May to October) and dry (November to April). For simplicity, we assume that demand is equally divided across all the time slices: wet-day, wet-night, dry-day, and dry-night.

South Sudan has very little existing infrastructure, including 30 MW of electricity generation capacity mainly from portable diesel generators (World Bank, 2013). Based on a planning report by Hatch (2014), we model a largely hypothetical electricity grid connecting 5 hydro power plants and 11 thermal power plants to meet electricity demand at 13 different demand centers located across 10 constitutionally established states. Electricity transmission links between demand centers and between demand centers and the proposed hydro and thermal plants are modeled explicitly, as shown in Fig. 1.

As an alternative to the proposed hydro and thermal plants, we also include distributed solar photovoltaics (PV), which can be built at each demand point as an alternative source of electricity generation (Fig. 1), but produce zero electricity at night. An advantage of distributed solar PV is its modularity; it can be deployed on a small scale at critical locations and built up over time. For example, Mozersky and Kammen (2018) suggest that solar PV could initially be funded by international donor governments and used to generate electricity on protected compounds associated with non-governmental organizations, UN agencies and peacekeeping bases, and protection of civilians (POC) camps. We omit consideration of centralized solar PV facilities, as they suffer from the same vulnerabilities as the centralized hydro and thermal plants. In this analysis, we also omit consideration of storage coupled to the solar PV systems. While storage is a feasible option that could allow solar generated electricity to meet demand at night, modeling it properly requires a higher temporal resolution of electricity supply and demand, ideally hourly, in order to capture the individual store and dispatch decisions. Such a representation is beyond the scope of this study.

Generator locations are based on Hatch (2014), and Table 2A of the appendix maps each existing and proposed generator to South Sudan's ten states.

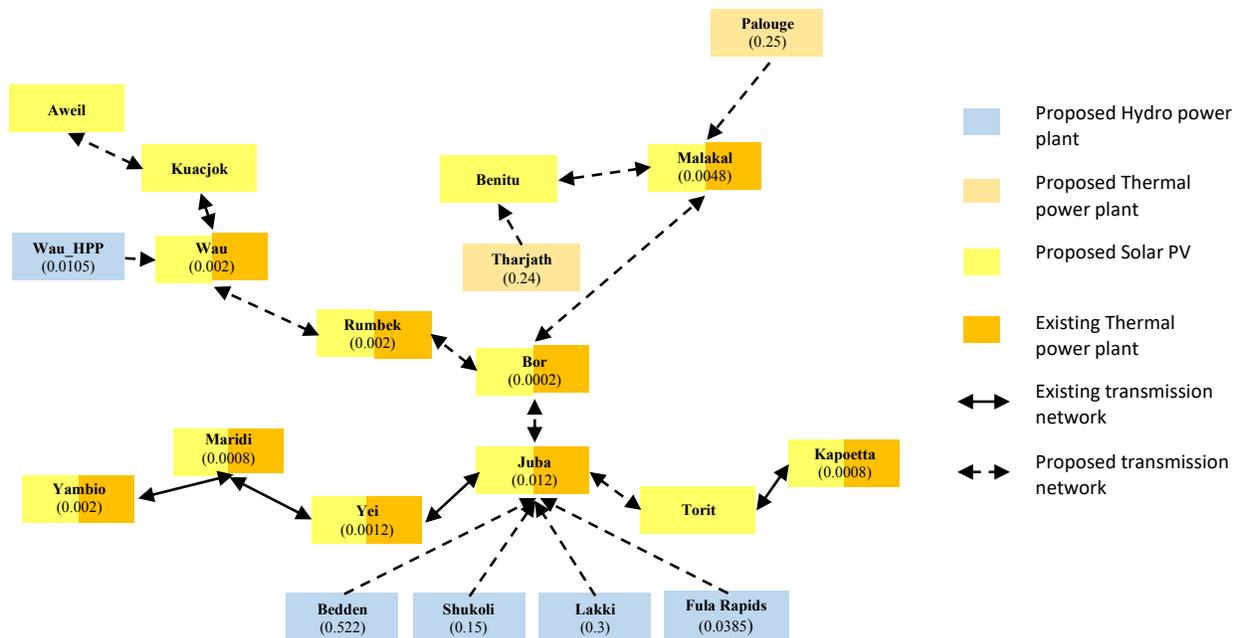

**Figure 1**. Layout of the modeled South Sudan system. Solar PV can be constructed within each of the 13 demand centers. Existing and proposed capacities of hydro and thermal units are denoted in parentheses in GW.

For simplicity, we assume that the investment cost, fixed cost, variable cost, capacity factor and efficiency of a technology remain unchanged over the model time horizon. Cost and performance associated with hydro power plants are taken from Hatch (2014). EIA (2013) is used to estimate solar PV cost, and the location-specific solar PV capacity factors are taken from IRENA (2018). Both the investment costs associated with proposed thermal power plants and the efficiency of existing thermal power plants are taken from Hatch (2014). The rest of the cost and performance

  4

coefficients are taken from Trüby, J. (2014) due to the unavailability of region-specific data. All of the thermal power plants are assumed to run on diesel. The most recent available estimate on the diesel price is $1.98/liter (World Bank, 2017). Operations and maintenance costs for the transmission lines are omitted, as we assume they are small relative to their investment cost. Investments in new hydro and thermal capacity (Fig. 1) also require dedicated transmission lines. The model ensures that the transmission lines are built along with the power plants. No new investment in thermal or hydro capacity other than the proposed capacity denoted in Fig. 1 is permitted. Because hydro capacity requires a significant lead time, we model a 5-year, 1-period delay between hydro capacity construction and when it can generate electricity. All future costs associated with technology deployment, operations, and incurred damage are discounted to the base year (2017) using a 3% global discount rate. For simplicity, all generating technologies are assumed to have a 30-year technical lifetime.

Electricity demand is an exogenous input taken from a comprehensive infrastructure action plan produced by the African Development Bank Group in 2013 (AfDB, 2013). The report includes a load forecast for the ten former South Sudan state capitals as well as for three additional important population centers. The demand growth in the short-term is estimated based on the historical demand growth trends of the South Sudan Electricity Corporation (SSEC). The demand forecast for the medium- to long-term is estimated based on projected consumption by customer and tariff categories, including domestic/household, commercial, and government. The potential demand in Juba and Malakal is based on field surveys undertaken by the SSEC. Commercial demand is assumed to grow at 10% per year while the government demand is projected to grow at 6% per year. Estimated demand by region as given in Hatch (2014) is provided in the Appendix.

Conflict will likely raise the electricity price, which in turn will suppress projected demand growth. Estimating demand elasticity in a country such as South Sudan is extremely difficult given the paucity of data, but we can assume that most of the population will be unwilling to tolerate high electricity prices. For simplicity, we assume that when electricity prices exceed a threshold value, consumers will choose to curtail their demand. We refer to this threshold as the curtailment cost. We assume that curtailment cost varies from 0.1 to 0.8 $/kWh, which encapsulates the range estimated by Oseni and Pollitt (2013). In addition, Steinbuks and Foster (2010) use the marginal cost method of revealed preference approach, and estimate the outage cost in sub-Saharan Africa between 0.13 – 0.76 $/kWh (2007 prices). Throughout our analysis, we vary the curtailment cost, which serves as a sensitivity on the level of consumer demand met by electricity supply. The required amount of electricity supply decreases as the prescribed curtailment cost decreases.

## 2.3 Method of Morris
Before conducting the stochastic optimization, we apply the Method of Morris (Morris, 1991) in order to identify the input parameters that produce the largest change on total system cost. The results are used to prioritize data collection needs and inform the stochastic program model formulation (Francesca, 2007). Unlike other sensitivity methods (Saltelli et al., 2004, 2005; Cacuci and Ionesco-Bujor, 2004, Pappenberger et al., 2006), the Method of Morris falls under the simplest class of one-factor-at-a-time (OAT) screening techniques. It assumes $l$ levels per input factor and generates a set of trajectories through the input space. As such, the Method of Morris generates a grid of uncertain model input parameters, $x_i, i = 1, \ldots k$, where the range $[x_i^-, x_i^+]$ of each uncertain input parameter $i$ is split into $l$ intervals of equal length. Each trajectory starts at different realizations of input parameters chosen at random and are built by successively selecting one of the inputs randomly and moving it to an adjacent level. These trajectories are used to estimate the mean and the standard deviation of each input parameter on total system cost. A high estimated mean indicates that the input parameter is important; a high estimated standard deviation indicates important interactions between that input parameter and other inputs.

In this analysis, we consider curtailment cost, electricity demand as well as generator capacity factor, fixed operations and maintenance costs, and investment costs as uncertain parameters. While the latter three parameters vary by generator type, they were grouped together in Method of Morris such that the same proportional perturbation to each parameter is made across each of the three generator technologies. For example, in a given Method of Morris iteration, a 3% perturbation to capacity factor is applied to solar PV, thermal, and hydro plants uniformly. This approach reduces the number of required trajectories and therefore the computational burden associated with Method of Morris. Given five model inputs, we have $l^5$ points in the grid, which we call the 'experimental space' $\mathbb{F}$. From $\mathbb{F}$, $r$ points are drawn at random, and the model is evaluated to obtain its objective function value at each of the $r$ points. For each model input value defined as $x_i, i = 1, \ldots, k$, the elementary effect of $i^{th}$ input factor on the objective function $F(x)$ is defined as

$$d_i(x) = \left( \frac{F(x_1, \ldots, x_{i-1}, x_i + \Delta, \ldots, x) - F(x)}{\Delta} \right) \quad (1)$$

where, $\Delta$ is a value such that the point $(x_1, \ldots, x_{i-1}, x_i + \Delta, \ldots, x_k)$ remains in the experimental space $\mathbb{F}$ for all $i, i = 1, \ldots, k$. Further, $\mu_i^*$ is the estimated mean of the distribution of elementary effects, $G_i$: $\mu_i^* = |d_i(x)| \sim G_i$ (Campolongo et al., 2007). It addresses the screening problem by identifying the subset of the model parameters which are not influential and hence can be fixed to any value within their ranges of uncertainty without significantly affecting the model outcome of interest. To conduct this part of the analysis, we utilized SALib (Herman and Usher, 2017), an open source Python library, which includes a complete implementation of the Method of Morris.

### 2.4 Stochastic Problem Formulation

Planners in countries such as South Sudan must make decisions in the face of deep uncertainty regarding future conflict. Energy system models often ignore future uncertainty by assuming perfect foresight across the entire model time horizon. In this case, individual model scenarios assume the future is known with certainty prior to the model run. In the case of conflict modeling, conventional scenario analysis would mean assuming a specific conflict scenario for a given model run. While such an approach using conventional scenario analysis can yield insight, it does not lead to a single unified strategy in the face of future uncertainty (Kann and Weyant, 2000). A key challenge for planners in FCV countries is to develop a near-term investment strategy that accounts for future conflict uncertainty.

To address this challenge, we frame the problem as a multi-stage stochastic optimization, which allows us to directly account for conflict uncertainty by incorporating it within the model formulation. Performing stochastic optimization requires us to consider future outcomes, assign probabilities to those outcomes, and quantify the effects of those outcomes. This information is organized in a scenario tree (Fig. 2), which describes the set of possible outcomes that may unfold over time[1]. Optimization is performed simultaneously over the entire scenario tree. Because the scenario tree accounts for different probability-weighted outcomes, the resultant model solution provides near-term planning strategy that account for future conflict-related uncertainty.

Conflict can result in many forms of damage within the power system, including damage to generators, transmission and distribution lines, fuel supply infrastructure and logistics, and maintenance. For simplicity, we focus on potential damage to the generators themselves. Thus we design a scenario tree that considers a binary outcome at each time stage: either generator damage occurs or it does not. Subjective probabilities denoting the probability of damage to generators during conflict are assigned to each branch in the scenario tree. Whether armed conflict leads to generator damage is deeply uncertain; and thus we create the scenario tree around this factor in order to explicitly test the effect of different damage probabilities on the deployment of new capacity. We assume that damage to generators results in increased fixed operations and maintenance (O&M) cost and a decrease in capacity factor for 5 years following the time of damage. We begin by describing the scenario tree structure, followed by the damage estimation method.

*2.4.1 Scenario Tree Structure*

The stochastic programming model includes three uncertain time stages and two branches (realizations) per node within the event tree. The two branches emanating from each node represent the possibility that generator damage either occurs with probability $Pr(D)$ or does not occur with probability $(1 - Pr(D))$ in the next time stage. To test the system response to uncertain damage probabilities, we assume two sets of nodal probabilities: high ($Pr(D) = 0.9$) and medium ($Pr(D) = 0.5$), as shown in Table 1. We also tested a low nodal ($Pr(D) = 0.1$) probability of damage but found results similar to the base case, and therefore omitted them in the results section. Considering conflict-related damage with a binary outcome over three uncertain time stages leads the scenario tree shown in Fig. 2.

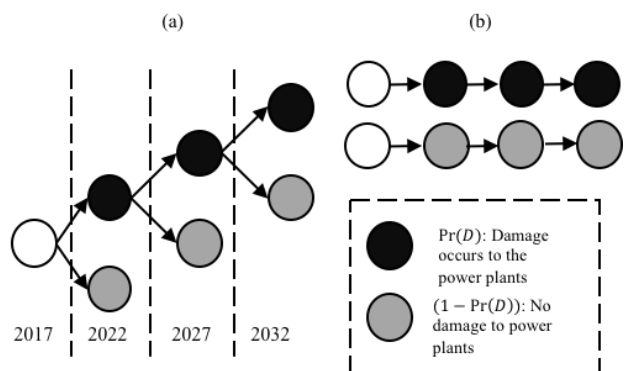

**Figure 2.** (a) Scenario tree representation with three uncertain stages, where each region of South Sudan has its own damage intensity and each demand center within a given region experiences the same damage intensity. (b) Given this tree structure, there are 8 potentially different pathways through the scenario tree representing different combinations of damage versus no damage to generators at each of the three uncertain time stages. Two of the 8 pathways are shown.

---

[1] For a formal mathematical treatment of stochastic optimization, see Dantzig, (2010), Shapiro et al. (2009), and Birge and Louveaux (2011).

*Table 1: Node probabilities of a scenario tree*

| Damage probability | High | Medium |
|---|---|---|
| **Probability of generator damage: *Pr(D)*** | 0.90 | 0.50 |
| **Probability of no generator damage: 1 – *Pr(D)*** | 0.10 | 0.50 |

*2.4.2 Damage Estimation*

With the scenario tree shown in Fig. 2, we must quantify the damage on the branches where conflict-related damage occurs. The resultant damage quantification is embedded within the scenario tree used to perform the stochastic optimization, and affects the investment decisions in different capacity types. In this analysis, we assume that conflict-related damage takes two forms: an increase in fixed operations and maintenance (O&M) cost, and a decrease in capacity factor. The increased cost and degraded capacity factor are both incurred for one model time period following the occurrence of damage. This section focuses on the damage estimation method; the numerical values are provided in the appendix. Notation used to estimate damage includes the following:

| | |
|---|---|
| $Pr(C)$ | Probability of conflict in South Sudan |
| $RCR$ | Regional conflict rate, which represents the percentage of conflict occurring in a specific region given that conflict occurs in South Sudan |
| $FOM$ | Fixed operations and maintenance cost ($/kW-yr) |
| $EFOM$ | Estimated increase in $FOM$ based on the investment cost, $RCR$, $Pr(C)$, and $DR$ |
| $DR$ | Damage rate representing the rate of $FOM$ increase |
| $CF$ | Capacity factor |
| $ECF$ | Estimated capacity factor following damage; based on $CF$, $RCR$, $Pr(C)$, and $DCF$ |
| $DCF$ | Percent reduction in $CF$ following damage |

Because conflict is likely to persist in South Sudan over the next two decades, we set the conflict probability, $Pr(C)$, at 0.9 in all scenarios. The regional conflict rate (RCR) is estimated from Raleigh et al. (2010), which has monitored the conflicts occurring in South Sudan since its independence in 2011. South Sudan is constitutionally divided into 10 regions. We assume that conflict occurs in each of these regions with frequency values based on conflicts in 2016 (see Table 3A in the appendix for RCR values). Generator damage, represented by an increase in FOM (denoted EFOM) and decrease in CF (denoted ECF), varies by generator type and the location of the generator.

Following Mozersky and Kammen (2018), we make the critical assumption that solar photovoltaics (PV) will be more resilient to conflict-related damage, and thus the changes to solar-related fixed O&M and capacity factor are less severe compared to hydro and thermal plants. In this analysis, the damage rate (DR) represents the increase in the fixed operations and maintenance cost (FOM) equivalent to the annual payment on investment cost. For thermal and hydro power plants, the DR is calculated using 100% of the annual payment on capital, while for solar PV, the DR is calculated using 10% of the annual payment on capital. For example, we assume an investment cost of 3350 $/kW for solar photovoltaics. Over a 30-year lifetime, the annual payment on capital is 171 $/kW-yr using a 3% discount rate. The damage is assumed to be 10% of this cost, or 17.1 $/kW-yr. Since the solar fixed O&M is 25 $/kW-yr (see Table 3A), the damage rate is calculated as (17.1 + 25) / 25 $/kW-yr, which equals 1.68. Thus damage to solar increases its fixed O&M by 68%. (See Appendix for detailed calculations of DR.)

We do not have data to derive an empirical value for the DR, and thus the assumed values here represent an informed judgment on our part. The difference in DR between solar versus hydro and thermal plants is intended to reflect the modularity and smaller scale of the solar installations. For example, 100 installations of 10 kW solar are likely to incur less damage than a single 1000 kW installation of thermal capacity.

While stochastic optimization explicitly addresses conflict uncertainty and how it could shape power system development, additional exogenous uncertainties are considered. First, as mentioned above, the subjective probability of conflict-related damage within a given model time period was tested at high (90%) and medium (50%) levels (Table 1). Second, given the high sensitivity to curtailment cost as illustrated in Fig. 1, the stochastic optimization is repeated at different curtailment cost values. Third, we test three different methods that translate the presence of conflict into power plant damage. The first damage estimation method calculates damage based on historical tallies of armed conflict by region within South Sudan (Raleigh et al., 2010). The second method is similar to the first, but the damage estimates are scaled up such that maximum damage is incurred by at least one generator of each type. The third method ignores differences in regional conflict frequencies, and maximizes damage estimates assigned to each individual generator. The three different damage estimation methods are described below.

*Regional Damage*
In this method, generator damage is proportional to past conflict frequency by region, such that power plants in more conflict-prone regions will have higher damage costs. In this case, we increase the base FOM by the product of *FOM*, the maximum damage rate, *DR*, the regional conflict rate, *RCR*, and the probability of conflict, $Pr(C)$. We use a similar formula to evaluate the reduced capacity factor, *ECF*. Most of the proposed hydro capacity is in Eastern and Central Equatoria, where a higher rate of conflict was observed in 2016. As a result, the model prefers to build relatively expensive distributed solar PV over the cheap hydro power to avoid the damage cost.

$$EFOM = FOM \times (1 + DR \times RCR \times Pr(C)) \quad (2)$$
$$ECF = CF \times (1 - (1 - DCF) \times RCR \times Pr(C)) \quad (3)$$

*Intensified Regional Damage*
In this method, a scaling factor $\alpha$ is added to Equation (2) and a scaling factor $\beta$ is added to Equation (3). The value of $\alpha$ is calibrated such that the annual damage cost incurred by at least one hydro and one thermal unit over a single model time period is equivalent to the annual payment on its capital cost. Likewise, the $\alpha$ value for solar is calibrated such that the annual damage cost associated with at least one solar installation is equivalent to 10% of the annual payment on its capital cost over one model time period. Similarly, the value of $\beta$ is calculated so that capacity factor of the same thermal and hydro power plants is decreased by 90% while the capacity factor of the same solar PV units is decreased by 10%. We note that the damage cost varies across individual generators because the *RCR* varies by region and the *DR* varies by plant type. This method leads to higher damage costs compared to the "regional damage method" above.

$$EFOM = FOM \times (1 + \alpha \times DR \times RCR \times Pr(C)) \quad (4)$$
$$ECF = CF \times (1 - \beta \times (1 - DCF) \times RCR \times Pr(C)) \quad (5)$$

*Max Damage*
This method produces the highest damage estimates. In this case, the *RCR* and *Pr(C)* terms are removed. For all hydro and thermal plants, the annual increase in *FOM* for each year in one time period is equal to the annual payment on its capital, and for all solar units, the annual increase is in FOM is equal to 10% of the annual payment on its capital. Similarly, for all years in a single model time period, the capacity factor of all the thermal and hydro power plants is decreased by 90%, and all solar PV capacity factors are decreased by 10%.

$$EFOM = FOM \times DR \tag{6}$$
$$ECF = CF \times (1 - DCF) \tag{7}$$

## 2.5 Metrics to assess value: EVPI, VSS and ECIC

Decision makers should be able to assess the economic value of plans made using stochastic programs. In this paper, we use the expected value of perfect information (EVPI) (Birge and Louveaux, 2011) and the value of the stochastic solution (VSS) (Birge, 1982) to characterize the economic impact of conflict damage on power systems and the economic value of the hedging strategy, respectively. In addition, we introduce a third metric called the 'expected cost of ignoring conflict' (ECIC) that estimates the savings associated with pursuing the stochastic programming solution rather than a least cost (naive) solution that ignores conflict completely. The resultant values associated with all three metrics vary depending on the ESOM parameterization and the scenario tree representation used in the stochastic optimization.

### 2.5.1 Expected Value of Perfect Information (EVPI)

The EVPI represents the amount of money that decision makers should be willing to pay in order to eliminate future uncertainty. Even when the EVPI is low, naïve decisions that ignore future uncertainty can perform poorly (Mercier and Van Hentenryck, 2007). Each forward path in the scenario tree is first solved deterministically and then the expected cost over those scenarios is calculated. This is known as the expected value of the *wait-and-see* solution (Madansky, 1960):

$$\mathbb{E}_\omega[Z_{DM}^\omega] = \sum_{\omega \in \Omega} p_\omega (Z_{DM}^\omega) \tag{8}$$

where $Z_{DM}^\omega$ is a deterministic model specified according to the set of forward paths $\Omega$, and $\omega$ represents a single scenario realization. The EVPI, which represents the difference between the wait-and-see and stochastic solutions, is then computed for multi-stage stochastic programs:

$$EVPI = Z_{RP} - \mathbb{E}_\omega[Z_{DM}^\omega] \tag{9}$$

where $Z_{RP}$ represents the multi-stage, stochastic program solved using the entire scenario tree instead of optimizing a single forward path.

### 2.5.2 Value of Stochastic Solution (VSS)

The VSS assesses the incremental value of the stochastic solution compared to a deterministic solve that considers the uncertain parameters represented at their expected values. The expected value of the uncertain parameters in the scenario tree is given as:

$$\bar{\xi}_t = \sum_{\omega \in \Omega} p_\omega (\xi_t^\omega) \tag{10}$$

where $\xi_t^\omega$ is the realization of the uncertain parameters in scenario $\omega$ at time stage $t$ which has a probability $p_\omega$ of occurrence. The deterministic model is specified for this purpose by considering the future realization of the uncertain parameter, $\xi_t$, for time period, $t = 2, \ldots, T$, at the expected value $\bar{\xi}_t$. We define $\bar{\xi} = [\bar{\xi}_2\ \bar{\xi}_3\ \cdots\ \bar{\xi}_t\ \cdots\ \bar{\xi}_T]$ and we represent this problem by $Z_{DM}(\bar{\xi})$. Figure 3 depicts a deterministic three-stage problem where the uncertain parameters are defined to be at their expected values for Stages 2 and 3 respectively.

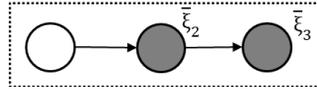

**Figure 3.** Single forward path in a three-stage problem used to solve $Z_{DM}(\bar{\xi})$

Results obtained in the first time stage of the deterministic model $Z_{DM}(\bar{\xi})$ are fixed as the first stage decisions for the stochastic program, and the optimization over the scenario tree is performed. Following Escudero et. al. (2007), we let $\bar{x}_t, \forall t \in T$, be the optimal solution at a given time stage obtained by solving the deterministic problem, $Z_{DM}(\bar{\xi})$. For calculating VSS, the decision vector of a recourse problem at Stage 1, $x_1$, is fixed as the optimal decision vector at Stage 1, $\bar{x}_1$, obtained by solving $Z_{DM}(\bar{\xi})$. If we let the solution to the recourse problem be denoted by $Z_{RP}(x_1 = \bar{x}_1)$, then the value of the stochastic solution can be defined as:

$$VSS = Z_{RP}(x_1 = \bar{x}_1) - Z_{RP} \tag{11}$$

If the VSS is small, it implies that the stochastic optimization conveys little value, since the future uncertainty can be well-represented by a deterministic formulation.

*2.5.3 Expected Cost of Ignoring Conflict (ECIC)*

The expected cost of ignoring conflict (ECIC) represents the savings associated with following the hedging strategy produced by the stochastic optimization instead of naively following a forward path that does not consider conflict and then having to take recourse action. ECIC is conceptually similar to a well-known metric, the expected cost of ignoring uncertainty (ECIU) (Birge and Louveaux, 2011), however, here we focus on only one particular naïve scenario in which conflict is ignored entirely. ECIC is also similar to the VSS, except the first stage decisions reflect the solution to the naïve scenario rather than the solution to deterministic scenario based on expected values for uncertain parameters.

We have two scenarios emanating from each node in the scenario tree: generator damage occurs ($\omega_1$) and no generator damage occurs ($\omega_2$). Hence, the deterministic model is represented as $Z_{DM}(\xi^{\omega_2})$, where $\xi^{\omega_2}$ represents the realization of uncertain parameters when no damage occurs across the planning horizon. Once the decision is made at Stage 1 using the naïve solution, we consider its cost in all the forward paths represented in the scenario tree. ECIC assesses the incremental value of a decision plan obtained using the recourse problem ($Z_{RP}$), where future uncertainty is explicitly considered instead of the naive solution that ignores it and requires significant recourse action in future periods.

We let $\tilde{x}_t, \forall t \in T,$ be the optimal solution at a given time stage obtained by solving the deterministic problem, $Z_{DM}(\xi^{\omega_2})$. To calculate ECIC, the decision vector of a recourse problem at Stage 1, $x_1$, is fixed at the optimal decision vector at Stage 1, $\tilde{x}_1$, obtained by solving the naïve scenario $Z_{DM}(\xi^{\omega_2})$. If we let the solution to this recourse problem be denoted by $Z_{RP}(x_1 = \widetilde{x_1})$, then the expected cost of ignoring conflict can be defined as:

$$ECIC = Z_{RP}(x_1 = \widetilde{x_1}) - Z_{RP} \qquad (12)$$

A small ECIC suggests that ignoring generator damage is inexpensive, and hence the stochastic solution does not yield much value.

## 3. Results and Discussion

We begin by describing the Method of Morris results, which inform the analysis performed with the stochastic version of the model. Next, results from the stochastic optimization are presented under different assumed conflict scenarios. We conclude by discussing the stochastic output metrics – EVPI, VSS and ECIC – and use them to draw insights about future electric power development in South Sudan.

### 3.1 Identifying key input sensitivities

Before conducting the stochastic optimization, we run a base case that serves as a benchmark and assumes no conflict in South Sudan. We apply Method of Morris to the base case in order to identify the input parameters that produce the largest effect on total electricity supply cost over the time horizon. This initial sensitivity analysis ignores conflict risk, which is addressed explicitly in the stochastic programming model. Key parameters tested include the electricity curtailment cost and end-use electricity demand as well as the capacity factors, investment costs, and fixed operations and maintenance costs associated with new electric generating units. Fig. 4 indicates that the cost of electricity supply is most sensitive to electricity curtailment costs.

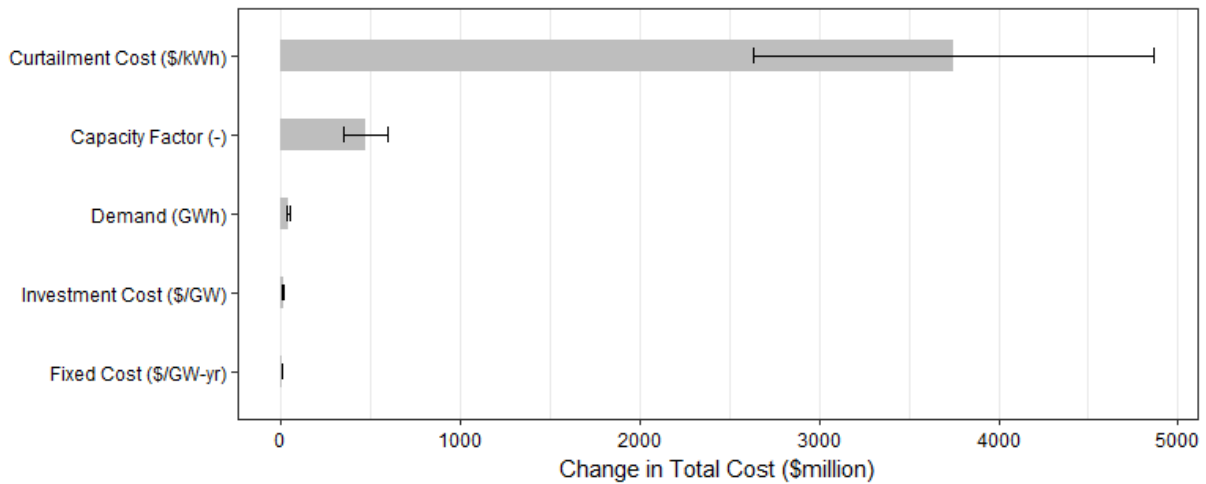

**Figure 4.** Change is the total cost of energy supply given a ±20% change in the value of five different input parameters using the Method of Morris. The length of the bars indicates the average effect of each parameter on total system-wide cost, and the error bars indicate the standard deviation across an ensemble of runs.

### 3.2 Capacity expansion under conflict uncertainty

Results from the first model time period (2017) reveal how the near-term hedging strategy produced by the stochastic optimization accounts for conflict uncertainty (Fig. 5). The amount of

    

installed capacity varies by the damage probability, curtailment cost, and damage estimation method. However, some patterns are evident. At all but the lowest curtailment value of 0.10 $/kWh, solar PV is a cost-effective option to meet demand given its greater resilience in the face of conflict. The combination of high damage probability and high damage values decreases the deployment of large hydro plants. In the case with high damage probability and maximum damage (Fig. 5a), it is most-effective to utilize solar PV and simply curtail demand at night when the curtailment cost is less than 0.6 $/kWh. Thermal capacity is only deployed when the curtailment cost is 0.6 $/kWh or above. Fig. 5 suggests that an explicit consideration of conflict can have a large effect on near-term electric sector planning.

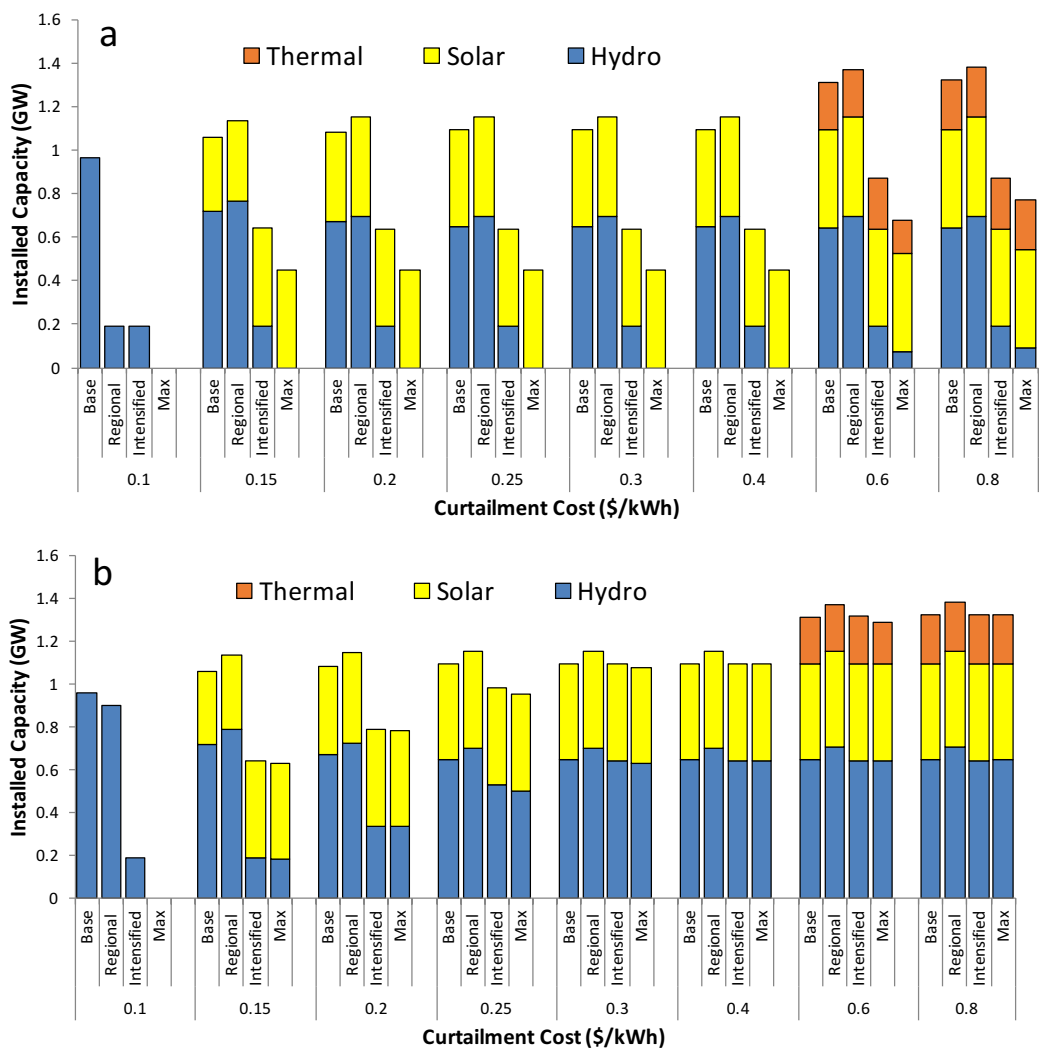

**Figure 5.** Installed capacity in the first model time period (2017) assuming (a) high probability of power plant damage, and (b) medium probability of power plant damage. The stochastic optimization is repeated for different curtailment costs and damage estimation methods: base (no damage), regional, intensified, and max damage. Differences in the total amount of installed capacity stem from differential use of demand curtailment and differences in technology-specific capacity factors. Given a 5-year delay in hydro availability, hydro constructed in 2017 is not available for generation until 2022.

The different conflict pathways represented in the scenario tree can lead to diverging deployment pathways over time. Fig. 6 illustrates how the deployment pathways of solar, hydro, and thermal differ under extreme scenarios; the vertical axis represents the difference in installed capacity between the scenario where generator damage occurs in each of the three future time periods, and the scenario where no generator damage occurs throughout the planning horizon. Under high damage probability (Fig. 6a), the expansion of hydro capacity is limited and there is little divergence between the no damage and all damage scenarios. As the expected damage increases (moving left-to-right in Fig. 6), there is greater divergence in the installed capacity by technology. Persistent conflict-related damage across the time horizon suggests the use of more solar PV and less hydro and thermal capacity. Differences in installed capacity between these extreme scenarios are larger under moderate damage probabilities because the expected damage costs are lower, allowing for greater variations in installed capacity as uncertainty about generator damage is revealed. This effect is amplified at higher curtailment costs because it is more cost-effective to build additional capacity than curtail demand. For example, under moderate damage probability and maximum damage, differences in 2032 installed capacity between different scenarios range up to 1 GW out of a total of 2.4 - 3.0 GW installed.

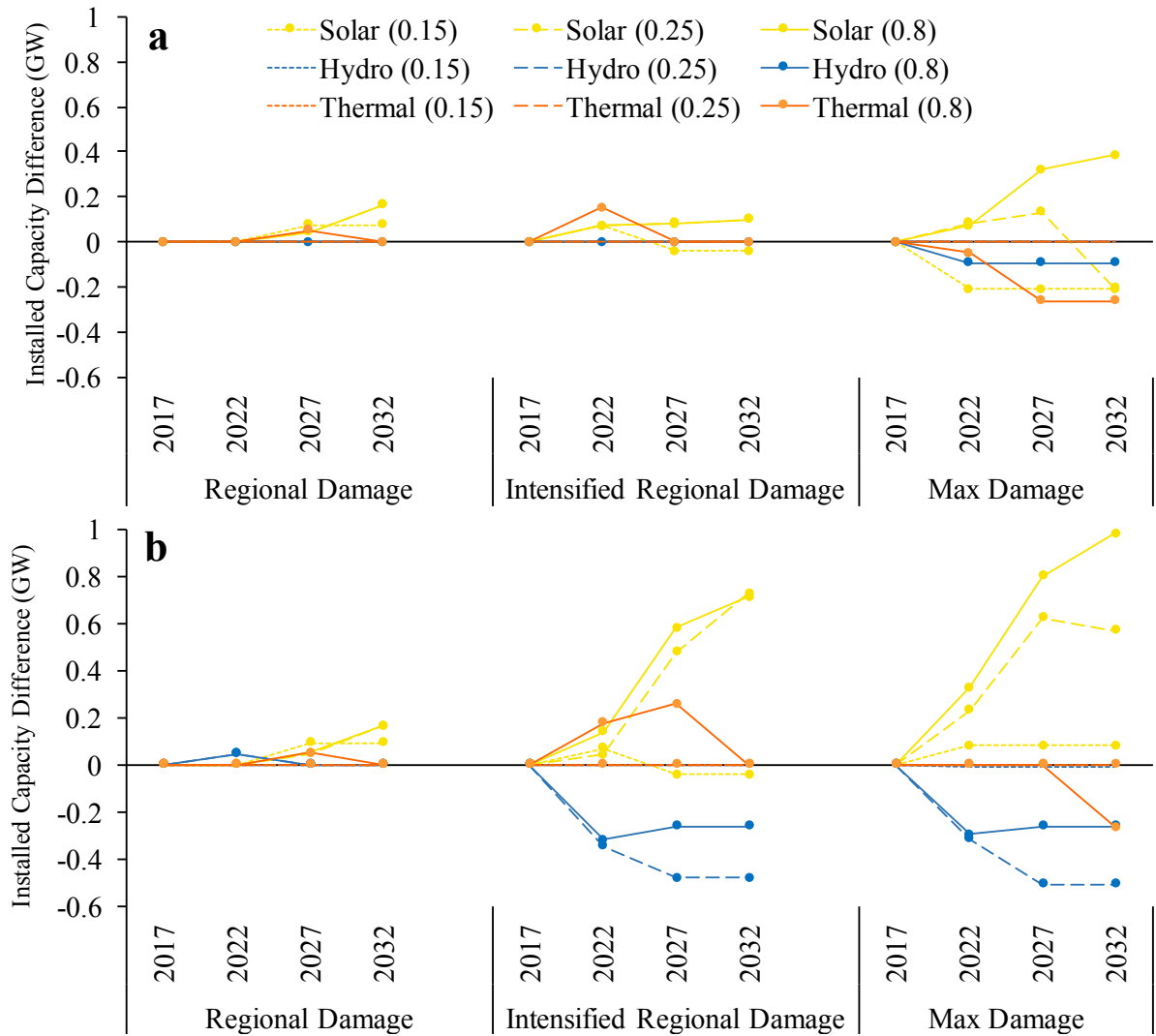

**Figure 6.** Difference in installed capacity between the two most extreme forward paths through the scenario tree: no generator damage and damage in every time period. Positive differences indicate higher installed capacity in the all-damage scenario. Differences are shown with (a) high damage probability and (b) medium damage probability, and in both cases, three different damage estimation methods. Differences are also shown at three different curtailment costs given in parentheses: 0.15, 0.25, and 0.8 $/kWh.

Fig. 7 presents the total cumulative installed capacity of hydro and solar by 2032 across all scenarios. Under low curtailment values of 0.15 $/kWh, only modest amounts of hydro and solar are deployed; it is more cost-effective to simply curtail demand. Under high damage probability (first and third rows of Fig. 7), only the scenario with regionalized damage costs and moderate to high curtailment costs allow for appreciable amounts of hydro capacity. In general, the high expected damages associated with high damage probabilities suppress the construction of hydro and increase the deployment of solar PV. By contrast, cumulative hydro capacity increases under medium damage probabilities (second row of Fig. 7). In this case, hydro remains cost-effective despite the anticipated damage costs.

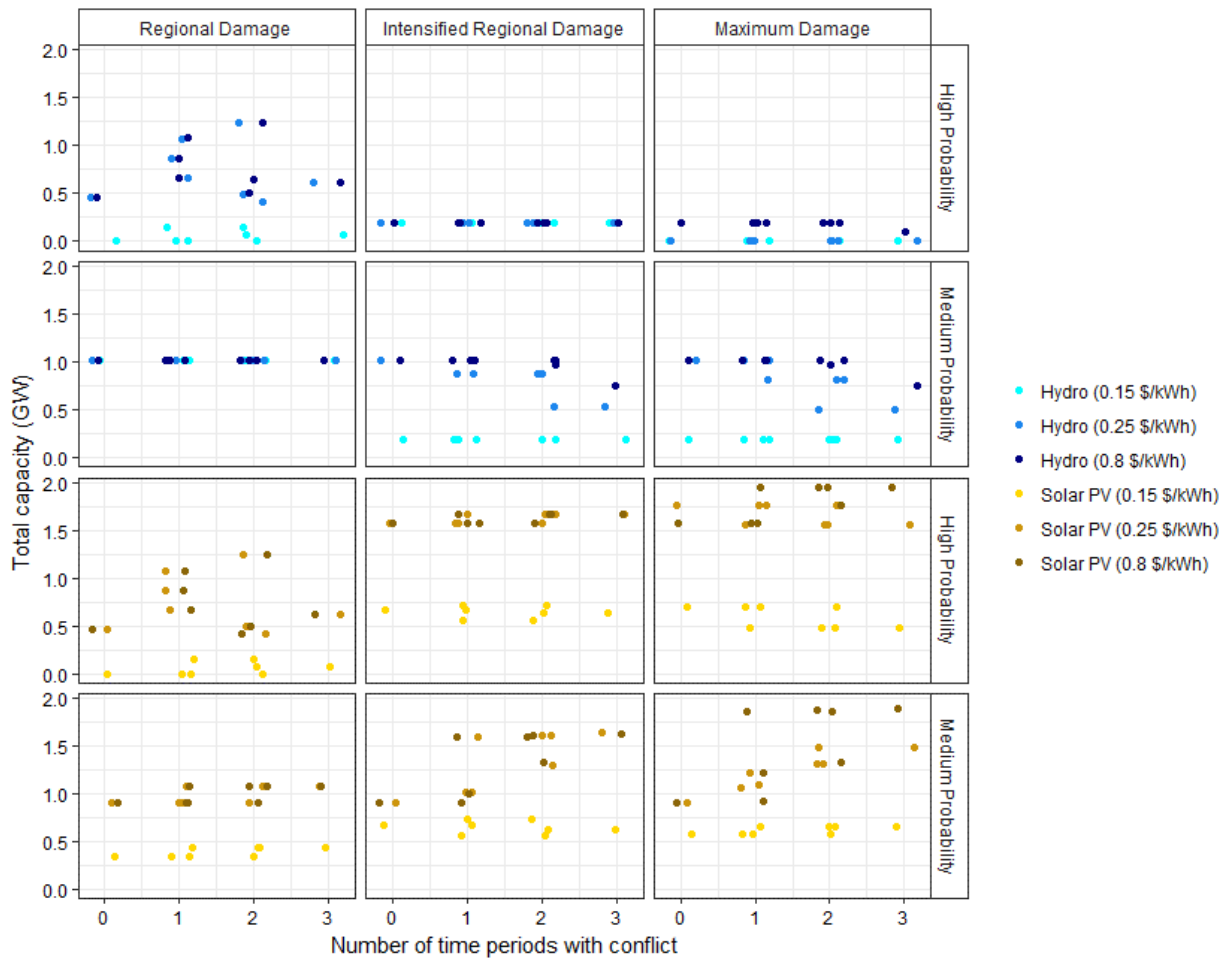

**Figure 7**. Cumulative installed capacity of hydro (top two rows) and solar (bottom two rows) as a function of the damage estimation method, damage probabilities (high or medium), and the number of time periods within the model time horizon that include conflict-related damage. Solar and hydro deployment are represented at different curtailment values, given in parentheses within the legend.

### 3.3 The value of hedging

Figs. 5-7 illustrate how assumptions about the probability of generator damage, method to estimate damages, curtailment cost, and number of time periods with conflict affect the installed capacity over time. It is also critical to assess how these factors affect the economics of electricity supply in South Sudan, and whether hedging strategies produced by the stochastic optimization provide economic value beyond a simpler deterministic model.

Fig. 8a presents the expected value of perfect information (EVPI), which can be interpreted in this context as the amount that South Sudan electricity planners should be willing to pay in order to eliminate uncertainty over future conflict. Fig. 8 plots the EVPI as a function of the exogenously prescribed curtailment cost, ranging from 0.1 to 0.8 $/kWh. The results indicate that the EVPI increases as the damage costs increase from the "regional damage" to "maximum damage" scenarios. In other words, as the economic impacts of conflict-related damage increase, the EVPI

 

increases. Under high probability of damage, the EVPI increases with curtailment cost. Higher curtailment costs imply that electricity is more highly valued, which means the model must rely more on electricity supply to meet demand at high curtailment costs. As a result, the value of information increases with curtailment cost under the high damage scenario, because the cost of damage can have a large effect on the total system cost. By contrast, under medium damage probability, the EVPI peaks at around 8% of the total baseline cost under the "intensified regional damage" and "maximum damage" scenarios at a curtailment cost value of 0.20 $/kWh. At higher curtailment values, the EVPI decreases because curtailment is effectively too expensive, and the strategy relies on building generating capacity, thus reducing the value of information. We also tested curtailment values exceeding 1 $/kWh in the high probability of damage case, and see the same decline in EVPI as shown in the medium probability case. The difference between the two scenarios is that the high damage probability makes the expected damage more costly, and thus the peak in EVPI occurs at a higher curtailment cost. Overall, the EVPI results indicate that when key input parameters take on intermediate values, the results are sensitive to how the uncertainty is resolved and therefore the value of information is comparatively high.

The value of the stochastic solution (VSS) in Fig. 8b follows a similar pattern to EVPI. The VSS indicates the relative value of the stochastic solution compared with a deterministic formulation that uses expected values for the uncertain damages. In the high damage probability case, the damages are high enough that the planning strategy is more straightforward and relies more on curtailment, which can be well-represented by a deterministic version. By contrast, the VSS results indicate that a medium damage probability and moderate curtailment values produce the largest VSS, reaching a peak of approximately 4% of the baseline cost. Thus, stochastic optimization has the greatest value in scenarios with intermediate values for damage probability and curtailment cost, where future uncertainty is highest.

The ECIC shown in Fig. 8c represents the difference between the hedging strategy and the naive solution that ignores conflict-related damage and requires recourse. Similar to the VSS, a small ECIC value implies that the hedging strategy produced by stochastic optimization, which accounts for future uncertainty in generator damage, conveys little economic advantage over a naïve least-cost pathway. The results indicate that the ECIC is at a maximum when the curtailment value is lowest (0.10 $/kWh), and decreases as the curtailment cost increases. Because the naïve solution does not consider conflict at all, a large amount of hydro capacity is built in 2017. However, conflict-related damage in later periods leads to expensive recourse that requires a shift toward solar. Recourse is more expensive under the high probability of damage case because the anticipated damages are higher. Because hydro includes a one-period delay in construction, more solar and thermal capacity is deployed in 2017 as the curtailment costs increase. Thus, higher curtailment costs force more solar deployment in the naïve scenario, which leads to less expensive recourse action in future periods when conflict occurs. Hence the decline in ECIC as a function of curtailment cost.

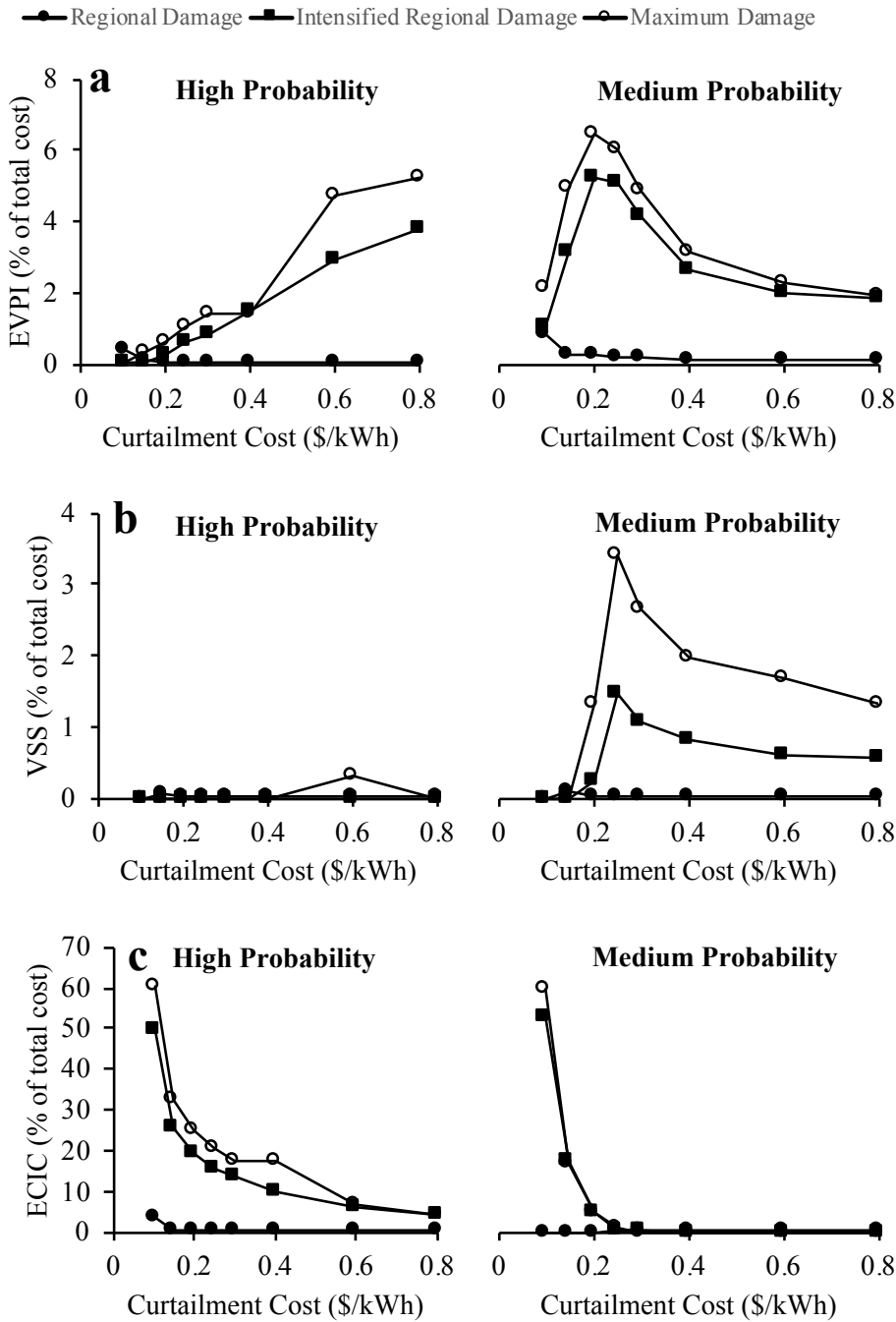

**Figure 8**. Measures of economic cost associated with future conflict uncertainty, including (a) the expected value of perfect information (EVPI), (b) value of the stochastic solution (VSS), and (c) the expected cost of ignoring conflict (ECIC). Methods producing higher damage estimates produce higher values for these metrics.

## 4. Conclusions

Fourteen of the top 20 most fragile countries are in sub-Saharan Africa (Fund for Peace, 2017). This modeling exercise demonstrates the need for such countries to explicitly consider the risk of conflict as they build out their electric power systems. We construct an analytical framework that employs an energy system optimization model along with sensitivity analysis and stochastic optimization to examine how potential future conflict can affect near-term electricity planning. We apply this framework to examine planning alternatives for South Sudan. We emphasize that the results presented here are meant to demonstrate how consideration of conflict in model-based analysis can inform planning, but additional refinement is required before it is used to inform decision making.

Even with the use of stochastic optimization, near-term deployment strategies still depend on exogenous assumptions; in this case, curtailment cost, damage probabilities, and the estimation method by which the presence of conflict can result in damage to electric generators. The ECIC results indicate that naively following a least cost planning strategy results in expensive recourse actions, particularly when the naïve scenario relies heavily on hydro. The ECIC peaks at approximately 60% of the baseline cost, which is extremely high because conflict-related damage converts much of the installed hydro capacity into a stranded asset. By comparison, van der Weijde and Hobbs (2012) examine transmission planning in Great Britain and estimate an overall ECIU value of 0.08% of the stochastic solution cost. Such high ECIC values in the South Sudan case suggest the potential for large economic losses in a country that can ill afford them.

Both the EVPI and VSS reach peak values under moderate damage probability and curtailment cost. In this intermediate range of input values, hedging based on consideration of future outcomes provides the most economic benefit. By contrast, more extreme scenario assumptions that make conflict-related generator damage either a near certainty or a remote possibility result in relatively straightforward decision strategies that do not require much hedging. Therefore, future work aimed at refining the range of input assumptions can help determine the utility of applying stochastic optimization to develop electric sector planning strategies under conflict.

5. **Caveats and Future Work**

As mentioned in the introduction, this analysis explores a simplified decision landscape that includes the explicit consideration of conflict in an electricity expansion planning exercise. Given the complexity of real world conflict dynamics, we made a number of simplifying assumptions.

First, we assume an exogenously specified demand increase over time, with demand curtailment occurring at a prescribed cost. The use of a curtailment cost – above which electricity is no longer demanded – is a simplifying assumption employed to make the model computationally tractable. Repeating the stochastic optimization at different curtailment cost values is functionally equivalent to adjusting the level of electricity demand that must be met. As the curtailment cost is lowered, less electricity demand will be met with supply. In reality, there would be a continuous response to prices: as prices increase, demand decreases, and vice versa. Future effort should focus on estimating the demand elasticity for electricity, which would allow the demand to adjust endogenously to realized electricity prices. Because South Sudan has never had significant electricity infrastructure, such elasticities would be highly uncertain, and could be incorporated as an uncertain parameter into the stochastic optimization.

Second, while we characterized three different methods to calculate damage to generators, more empirical data that quantifies how conflict affects electricity infrastructure would improve our

 19

analysis. We assume, for example, that the probability of conflict, *Pr(C)*, and regional conflict rate, *RCR*, do not vary with time. In addition, we assume independence among key damage-related parameters, including the probability of damage, *Pr(D)*, regional conflict rate, *RCR*, probability of conflict, *Pr(C)*, and the damage rate, *DR*. In reality, we would expect, for example, that the regional conflict would affect the damage rate, which would in turn affect the probability of damage. We are not aware of an existing dataset that would allow us to derive the correlation between these parameters. Future work could include additional sensitivity analysis to quantify how parameter correlation affects the damage estimates. The importance of time dependence and correlation among parameters also depends on the timeframe for analysis and the characteristics of the conflict. For example, given the persistence of the South Sudan conflict, assuming the conflict dynamics remain largely the same over the next 15 years may result in a plausible capacity expansion plan under conflict.

Also, in addition to considering damage to generators, future analysis should also consider damage to transmission and distribution lines, fuel supply infrastructure, and the resultant effect on cost and delay associated with maintenance. Damage to transmission and distribution infrastructure would disproportionately affect the centralized hydro and thermal plants, which typically rely on long distance, high voltage transmission lines to deliver electricity. By contrast, the modularity of solar allows for separate installations or microgids with a limited amount of distribution infrastructure. While this vulnerability is partially captured by the higher damage rate for hydro and thermal facilities compared to solar, the effect on transmission and distribution lines could be modeled explicitly. The vulnerability of fuel supply infrastructure would only affect the thermal plants, which only see limited deployment in the current analysis. Consideration of maintenance and repair should also be considered in future work. The effect on maintenance should consider plant locations, the ability to move through the terrain and source parts, and human capital required to repair and maintain the facilities. Given the modularity of solar and the ability to place it strategically within currently protected locations, we speculate that maintenance would incur larger costs and delays on the centralized plants.

# Appendix A: Supplementary Information

**Technology representation**

The base system includes the electric generating units and transmission lines shown in Table 1A. The base case assumes no conflict in South Sudan. All thermal power plants run on diesel. For simplicity, we assume that the investment cost, fixed and variable operations and maintenance costs, capacity factor, and efficiency of a technology remain unchanged for the entire time horizon. We also assume that fixed and variable cost for transmission lines and the variable cost for solar PV are negligible compared to their investment cost and therefore can be ignored. Moreover, solar PV cannot generate electricity at night and is thus assigned a capacity factor of zero at night. Location and status (new/existing) of the power plants, proposed transmission grid, all cost coefficients for proposed hydro power plants, investment and variable cost for thermal power plants, capacity factor and efficiency of hydro power plants, thermal power plants and transmission lines, and investment cost of transmission lines are taken from HATCH (2014). Fixed operations and maintenance costs for solar PV are considered 15% higher than given in EIA (2017). The investment cost of all solar PV is assumed to be 3350 $M/GW. This value is specific to Africa and is the average of the range of total installed cost (1820 - 4880 $/kW) (IRENA, 2015). Along with the equipment cost, it depends on the maturity of the domestic market, local labor, manufacturing cost, incentive levels, and structure. Data for location-specific capacity factors of solar PV are obtained from average annual direct solar irradiation taken from (Solargis, 2017). The most recent data available for the fuel price for thermal generators is $1.98/liter (World Bank, 2017).

**Table 1A.** Technology cost and performance assumptions.

| Location of the proposed power plants | Technology Name [a] | New or Existing (N/E) | Starting year | Variable cost ($M/GWh) | Fixed cost ($M/GWyr) | Investment cost ($M/GW) | Capacity Factor | Efficiency |
|---|---|---|---|---|---|---|---|---|
| Juba | SO1 | N | 2017 | 0 | 25 | 3350 | 0.3424 | 0.2 |
| Yambio | SO2 | N | 2017 | 0 | 25 | 3350 | 0.3196 | 0.2 |
| Maridi | SO3 | N | 2017 | 0 | 25 | 3350 | 0.3196 | 0.2 |
| Yei | SO4 | N | 2017 | 0 | 25 | 3350 | 0.3196 | 0.2 |
| Kapoeta | SO5 | N | 2017 | 0 | 25 | 3350 | 0.3424 | 0.2 |
| Torit | SO6 | N | 2017 | 0 | 25 | 3350 | 0.3424 | 0.2 |
| Benitu | SO7 | N | 2017 | 0 | 25 | 3350 | 0.3881 | 0.2 |
| Malakal | SO8 | N | 2017 | 0 | 25 | 3350 | 0.3424 | 0.2 |
| Bor | SO9 | N | 2017 | 0 | 25 | 3350 | 0.3196 | 0.2 |
| Rumbek | SO10 | N | 2017 | 0 | 25 | 3350 | 0.3424 | 0.2 |
| Wau | SO11 | N | 2017 | 0 | 25 | 3350 | 0.3881 | 0.2 |
| Kuajok | SO12 | N | 2017 | 0 | 25 | 3350 | 0.3881 | 0.2 |
| Aweil | SO13 | N | 2017 | 0 | 25 | 3350 | 0.4109 | 0.2 |
| Juba | TH1 | E | 2016 | 0.0464 | 20 | 0 | 0.75 | 0.42 |
| Yambio | TH2 | E | 2016 | 0.0464 | 20 | 0 | 0.75 | 0.36 |
| Maridi | TH3 | E | 2016 | 0.0464 | 20 | 0 | 0.75 | 0.34 |
| Yei | TH4 | E | 2016 | 0.0464 | 20 | 0 | 0.75 | 0.35 |
| Kapoeta | TH5 | E | 2016 | 0.0464 | 20 | 0 | 0.75 | 0.34 |
| Tharjath | TH6 | N | 2019 | 0.0399 | 20 | 1500 | 0.75 | 0.46 |
| Malakal | TH7 | E | 2016 | 0.0464 | 20 | 0 | 0.75 | 0.37 |
| Bor | TH8 | E | 2016 | 0.0464 | 20 | 0 | 0.75 | 0.36 |
| Rumbek | TH9 | E | 2016 | 0.0464 | 20 | 0 | 0.75 | 0.35 |
| Wau | TH10 | E | 2016 | 0.0464 | 20 | 0 | 0.75 | 0.34 |
| Palogue | TH11 | N | 2019 | 0.0399 | 20 | 1500 | 0.75 | 0.48 |
| Bedden | HY1 | N | 2031 | 0.0045 | 20 | 2500 | 0.3 | 0.9 |
| Shukoli | HY2 | N | 2031 | 0.0070 | 20 | 1800 | 0.32 | 0.9 |
| Lakki | HY3 | N | 2031 | 0.0022 | 20 | 2200 | 0.28 | 0.9 |
| Fula Rapids | HY4 | N | 2019 | 0.0004 | 20 | 3700 | 0.65 | 0.9 |
| Wau_HPP | HY5 | N | 2019 | 0.0004 | 20 | 13100 | 0.33 | 0.9 |
| **Bedden-Juba** | trans_Be_J | N | 2017 | 0 | 0 | 7.8 | 1 | 0.95 |
| **Shukoli-Juba** | trans_S_J | N | 2017 | 0 | 0 | 49.4 | 1 | 0.95 |
| **Lakki-Juba** | trans_L_J | N | 2017 | 0 | 0 | 30 | 1 | 0.95 |
| **Fula rapids-Juba** | trans_F_J | N | 2017 | 0 | 0 | 26.4 | 1 | 0.95 |
| **Juba-Yei** | trans_J_Y | E | 2016 | 0 | 0 | 0 | 1 | 0.95 |
| **Juba-Torit** | trans_J_T | N | 2017 | 0 | 0 | 37.9 | 1 | 0.95 |
| **Juba-Bor** | trans_J_B | N | 2017 | 0 | 0 | 59.6 | 1 | 0.95 |
| **Yei-Maridi** | trans_Y_Ma | E | 2016 | 0 | 0 | 0 | 1 | 0.95 |
| **Maridi-Yambio** | trans_Ma_Ym | E | 2016 | 0 | 0 | 0 | 1 | 0.95 |
| **Kapoeta-Torit** | trans_K_T | E | 2016 | 0 | 0 | 0 | 1 | 0.95 |
| **Bor-Malakal** | trans_B_M | N | 2017 | 0 | 0 | 91.3 | 1 | 0.95 |
| **Malakal-Benitu** | trans_M_Bu | N | 2017 | 0 | 0 | 42.2 | 1 | 0.95 |
| **Tharjath-Benitu** | trans_Th_Bu | E | 2016 | 0 | 0 | 0 | 1 | 0.95 |
| **Bor-Rumbek** | trans_B_R | N | 2017 | 0 | 0 | 86.1 | 1 | 0.95 |
| **Rumbek-Wau** | trans_R_W | N | 2017 | 0 | 0 | 82.6 | 1 | 0.95 |
| **Wau-Kuajok** | trans_W_Kj | N | 2017 | 0 | 0 | 23.1 | 1 | 0.95 |
| **Kuajok-Aweil** | trans_Kj_A | N | 2017 | 0 | 0 | 16.2 | 1 | 0.95 |
| **Palogue-Malakal** | trans_P_M | N | 2017 | 0 | 0 | 16.2 | 1 | 0.95 |
| **Wau_HPP-Wau** | trans_Wa_W | N | 2017 | 0 | 0 | 16.2 | 1 | 0.95 |

[a] 'SO' indicates solar PV, 'TH' indicates thermal plants running on diesel, 'HY' indicates hydro, and 'trans' indicates transmission lines.

Table 2A represents the region where each thermal and hydro power plant is located. For each region listed in Table 2A, there are one or more demand centers. The model allows one solar PV installation at each of the 13 demand centers. The electric power system under consideration also has 9 existing and 2 proposed thermal power plants as well as 5 proposed hydro power plants. Note that two of the biggest proposed hydro power plants (1 and 4) are located in Central Equatoria where the highest numbers of conflicts were observed in 2016.

**Table 2A** Location of all the generators by region.

| Region | Thermal power plant (TH) | | | | | | | | | | | Solar PV (SO) | | | | | | | | | | | | | Hydro power plant (HY) | | | | |
|---|---|---|---|---|---|---|---|---|---|---|---|---|---|---|---|---|---|---|---|---|---|---|---|---|---|---|---|---|---|
| | 1 | 2 | 3 | 4 | 5 | 6 | 7 | 8 | 9 | 10 | 11 | 1 | 2 | 3 | 4 | 5 | 6 | 7 | 8 | 9 | 10 | 11 | 12 | 13 | 1 | 2 | 3 | 4 | 5 |
| Central Equatoria | ■ | | | ■ | | | | | | | | ■ | | | ■ | | | | | | | | | | ■ | ■ | | | |
| Eastern Equatoria | | | | | ■ | | | | | | | | | | | ■ | ■ | | | | | | | | | | | ■ | |
| Jonglei | | | | | | | ■ | | | | | | | | | | | | | ■ | | | | | | | | | |
| Lakes | | | | | | ■ | | | | | | | | | | | | | | | ■ | | | | | | | | |
| Northern Bahr-el-Ghazal | | | | | | | | | | | | | | | | | | | | | | | ■ | | | | | | |
| Unity | | | | | | | | ■ | | | | | | | | | | ■ | | | | | | | | | | | |
| Upper Nile | | | | | | | | ■ | | ■ | | | | | | | | | | | | | | | | | | | |
| Warap | | | | | | | | | | | | | | | | | | | | | | ■ | | | | | | | |
| Western Bahr-el-Ghazal | | | | | | | | | ■ | | | | | | | | | | | | | ■ | | | | | | | ■ |
| Western Equatoria | | ■ | ■ | | | | | | | | | | ■ | ■ | | | | | | | | | | | | | | | |

### Electricity demand
Given the new country's challenging situation, there has not been a formal South Sudan Master Plan since its independence in 2011. Electricity demand is an exogenous input taken from a comprehensive infrastructure action plan produced by the African Development Bank Group in 2013 for South Sudan (AfDB, 2013). Electricity demand for each of the 13 demand locations considered for this analysis is given in Figure 1A.

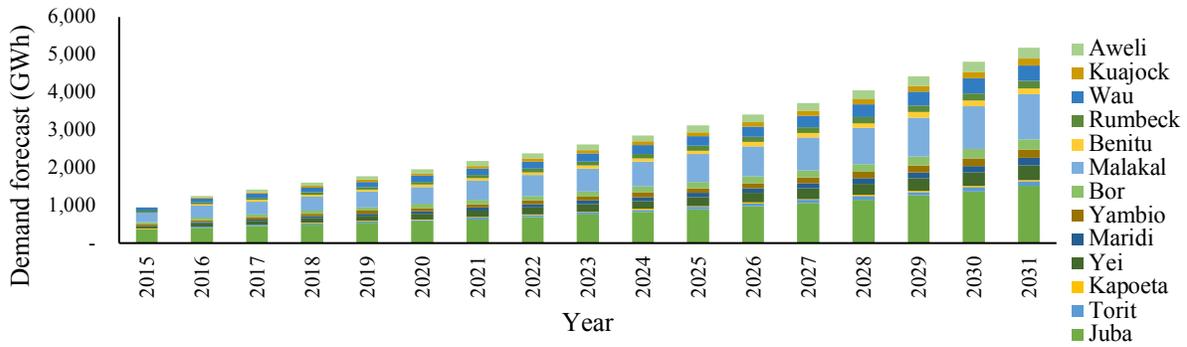

**Figure 1A:** Electricity demand forecast for South Sudan, drawn from AfDB (2013).

**Damage value estimation**

*Regional Conflict Rate*

The regional conflict rate (RCR) is estimated using the ACLED database (Raleigh et al., 2010), which has monitored the conflicts occurring in South Sudan since its independence in 2011. South Sudan is constitutionally divided into 10 regions. We assume that conflict occurs in each of these regions with the frequency values reported in Table 3A. The frequency represents the number of conflict incidences recorded in a given region during year 2016.

**Table 3A.** Regional conflict rate (RCR) calculated from frequency of conflict incidences in 2016 for 10 states in South Sudan (Raleigh et al., 2010).

| Regions | Abbreviation | Regional Conflict Rate | Frequency |
|---|---|---|---|
| Central Equatoria | CE | 0.298 | 268 |
| Eastern Equatoria | EE | 0.0855 | 77 |
| Jonglei | J | 0.0989 | 89 |
| Lakes | L | 0.0411 | 37 |
| Northern Bahr-el-Ghazal | NB | 0.0311 | 28 |
| Unity | U | 0.106 | 95 |
| Upper Nile | UN | 0.0767 | 69 |
| Warap | W | 0.0378 | 34 |
| Western Bahr-el-Ghazal | WB | 0.123 | 111 |
| Western Equatoria | WE | 0.102 | 92 |

*Damage Rate*

For hydro and thermal power plants, the damage rate (DR) represents the fractional increase in the fixed operations and maintenance cost (FOM) equivalent to the annual payment on investment cost. The increase in FOM or decrease in CF remains in effect for 5-year time period. The investment cost (IC) is amortized over 30 years with a 3% discount rate ($r$). The following example illustrates the calculation of DR for a thermal power plant, $DR_{TH}$, with an investment cost (IC) of $1500 million/GW and FOM of $20 million/GW-yr:

$$DR = \frac{\left(FOM + IC\left(\frac{r}{(1-(1+r)^{-p})}\right)\right)}{FOM}$$

$$DR_{TH} = \frac{\left(20 + 1500\left(\frac{0.03}{(1-(1+0.03)^{-30})}\right)\right)}{20}$$

$$DR_{TH} = 4.83$$

Similarly, due to the distributed nature of solar PV, we assume that $DR_{SO}$ is the rate of increase in FOM equivalent to 10% of the annual payment on investment. For solar PV with an investment cost (IC) of $3350 million/GW and FOM of $25 million/GW-yr, $DR_{SO}$ is calculated as follows:

$$DR_{SO} = \frac{\left(25 + 0.1 \times 3550\left(\frac{0.03}{(1-(1+0.03)^{-30})}\right)\right)}{25}$$

$$DR_{TH} = 1.68$$

In the event of damage to generators, Table 4A shows the estimated increase in FOM represented as the equivalent increase in investment cost and decrease in CF for each power plant, calculated using the three methods: Regional Damage, Intensified Regional Damage, and Maximum Damage. TH1, TH4, SO1, SO4, HY1 and HY3 are located in Central Equatoria where maximum conflict incidences are recorded in 2016. Hence, these power plants observe the highest increase in FOM and decrease in CF during conflict.

**Table 4A.** *EFOM* represents percent equivalent increase in investment cost while *ECF* represents equivalent percent of capacity factor remaining for each power plant after damage. *EFOM* and *ECF* for Regional damage ≤ Intensified regional damage ≤ Maximum damage. 'SO' indicates solar PV, 'TH' indicates thermal plants running on diesel, 'HY' indicates hydro power plant.

| Power plant[a] | Region | Equivalent capital cost increase | | | Capacity factor change | | |
|---|---|---|---|---|---|---|---|
| | | Regional Damage | Intensified Regional Damage | Maximum damage | Regional Damage | Intensified Regional Damage | Maximum damage |
| TH1 | Central Equatoria | 0.31 | 1.00 | 1.00 | 0.78 | 0.10 | 0.10 |
| TH2 | Western Equatoria | 0.09 | 0.28 | 1.00 | 0.94 | 0.75 | 0.10 |
| TH3 | Western Equatoria | 0.09 | 0.28 | 1.00 | 0.94 | 0.75 | 0.10 |
| TH4 | Central Equatoria | 0.31 | 1.00 | 1.00 | 0.78 | 0.10 | 0.10 |
| TH5 | Eastern Equatoria | 0.09 | 0.31 | 1.00 | 0.93 | 0.72 | 0.10 |
| TH6 | Unity | 0.08 | 0.26 | 1.00 | 0.94 | 0.77 | 0.10 |
| TH7 | Upper Nile | 0.08 | 0.27 | 1.00 | 0.94 | 0.76 | 0.10 |
| TH8 | Jonglei | 0.17 | 0.55 | 1.00 | 0.88 | 0.51 | 0.10 |
| TH9 | Lakes | 0.07 | 0.24 | 1.00 | 0.95 | 0.79 | 0.10 |
| TH10 | Western Bahr-el-Ghazal | 0.13 | 0.41 | 1.00 | 0.91 | 0.63 | 0.10 |
| TH11 | Upper Nile | 0.08 | 0.27 | 1.00 | 0.94 | 0.76 | 0.10 |
| HY1 | Central Equatoria | 0.25 | 0.97 | 1.00 | 0.78 | 0.10 | 0.10 |
| HY2 | Western Equatoria | 0.08 | 0.30 | 1.00 | 0.93 | 0.72 | 0.10 |
| HY3 | Central Equatoria | 0.26 | 1.00 | 1.00 | 0.78 | 0.10 | 0.10 |
| HY4 | Eastern Equatoria | 0.09 | 0.35 | 1.00 | 0.93 | 0.72 | 0.10 |
| HY5 | Western Bahr-el-Ghazal | 0.12 | 0.48 | 1.00 | 0.91 | 0.63 | 0.10 |
| SO1 | Central Equatoria | 0.06 | 0.10 | 0.10 | 0.98 | 0.90 | 0.90 |
| SO2 | Western Equatoria | 0.02 | 0.03 | 0.10 | 0.99 | 0.97 | 0.90 |
| SO3 | Western Equatoria | 0.02 | 0.03 | 0.10 | 0.99 | 0.97 | 0.90 |
| SO4 | Central Equatoria | 0.06 | 0.10 | 0.10 | 0.98 | 0.90 | 0.90 |
| SO5 | Eastern Equatoria | 0.02 | 0.03 | 0.10 | 0.99 | 0.97 | 0.90 |
| SO6 | Eastern Equatoria | 0.02 | 0.03 | 0.10 | 0.99 | 0.97 | 0.90 |
| SO7 | Unity | 0.02 | 0.03 | 0.10 | 0.99 | 0.97 | 0.90 |
| SO8 | Upper Nile | 0.02 | 0.03 | 0.10 | 0.99 | 0.97 | 0.90 |
| SO9 | Jonglei | 0.03 | 0.05 | 0.10 | 0.99 | 0.95 | 0.90 |
| SO10 | Lakes | 0.01 | 0.02 | 0.10 | 0.99 | 0.98 | 0.90 |
| SO11 | Western Bahr-el-Ghazal | 0.02 | 0.04 | 0.10 | 0.99 | 0.96 | 0.90 |
| SO12 | Warap | 0.01 | 0.02 | 0.10 | 1.00 | 0.98 | 0.90 |
| SO13 | Northern Bahr-el-Ghazal | 0.01 | 0.02 | 0.10 | 1.00 | 0.98 | 0.90 |